\documentclass[
  ,draft            
  ]
  {aipproc}

\layoutstyle{8x11single}

\begin{document}

\title{Evolution of Binaries in Dense Stellar Systems}


\author{Natalia Ivanova}{
  address={University of Alberta, Dept.\ of Physics, 11322-89 Ave, Edmonton, AB, T6G 2E7, Canada}
}

\begin{abstract}
In contrast to the field, the binaries in dense stellar systems are frequently
not primordial, and could be either dynamically formed 
or significantly altered from their primordial states.
Destruction and formation of binaries occur in parallel all the time.
The destruction, which constantly removes soft binaries from a binary pool,
works as an energy sink and could be a reason for cluster entering the binary-burning phase.
The true binary fraction is greater than observed,
as a result, the observable binary fraction evolves differently from the predictions.
Combined measurements of binary fractions in globular clusters suggest
that most of the clusters are still core-contracting.
The formation, on other hand, affects most the more evolutionary advanced stars, which 
significantly enhances the population of X-ray sources in globular clusters.
The formation of binaries with a compact objects
proceeds mainly through physical collisions, binary-binary and single-binary encounters;
however, it is the dynamical formation of triples and multiple encounters that principally 
determine whether the formed binary will become an X-ray source.
\end{abstract}

\maketitle


\section{Binaries evolution: destruction}

For low density environments, like solar neighborhood and for solar-type stars, the binary fraction is known to be $\sim 50\%$ 
 (e.g., \cite{1991A&A...248..485D}). Similarly, open clusters show 
relatively large measured binary fractions; as large as 70\% in some open clusters;
a clear anti-correlation with the age of the open cluster has been found \citep{2010MNRAS.401..577S}.
Most globular clusters have typically significantly lower measured binary fraction than open clusters,
although for some it is comparable to that of open clusters -- e.g., in
a core-collapsed NGC 6752, the upper limit for the binary fraction can be as high as 38\% \citep{1997ApJ...474..701R}
(although see also \cite{2010ApJ...709.1183M});
in a young  and sparse globular cluster in Ter 7, 
the binary fraction is estimated to be 51\% \citep{2007MNRAS.380..781S}.
For globular clusters, the measured binary fraction was found to anti-correlate 
with the total cluster luminosity \citep{2008MmSAI..79..623M};
the correlation with the cluster collisional parameter was found to be only marginal.
All together, measurements of binary fractions in open and globular clusters suggest that binaries
in clusters deplete with time and it happens more efficiently in more massive clusters.

There is no theoretical study that would suggest that the initial binary fraction should be significantly different
at the moment of star formation between proto-open and proto-globular clusters
\footnote{Recent studies showed that there could be more than one period of star formation in globular clusters,
and different generations could have been formed with different binary fractions \citep{2010ApJ...719L.213D}.}.
The consistency between the initially large binary fraction and currently measured low binary fraction
can be explained by the efficient binary destruction that continues throughout the entire cluster evolution.

Indeed, let us consider a population of primordial binaries with the initial periods $P=0.1-10^7$
and flat mass ratio distribution, evolved for 10 Gyr using StarTrack \citep{Bel08}. 
By this age, significant fraction of initial binaries is destroyed through evolution, mainly due to mergers,
or, in case of massive stars, in supernovae (see Fig.~1).
If we define the hardness of a binary $\eta$  as the ratio of the binary binding energy to
the kinetic energy that a $0.5 M_\odot$ object moving with $v=\sqrt{3}\times 10$ km/s has
(this energy is about an average kinetic energy of an object in a typical dense cluster and is referred here to as $kT$),
then it can be seen that by the age of 10 Gyr, almost all the hardest ($\eta \ge 100kT$) binaries are destroyed
via evolution.
Now we place a population of these binaries -- with primary masses drawn in accordance to initial mass function (IMF) from \citep{2002Sci...295...82K},
and with their initial eccentricities distributed thermally --
in a confined volume and keep the concentration of objects 
constant $n_{\rm c}=10^5$ pc$^{-3}$, constant velocity dispersion of 10 km/s and disallow
the objects to escape form this test volume.
If we let the same stars evolve for the same 10 Gyr, but allow them to interact dynamically, 
we find that - not surprisingly  - almost all the soft binaries ($\eta\ge 1kT$) are destroyed 
by dynamical encounters, except those with the collision time\footnote{this is the time-scale for a binary 
to undergo a strong encounter with another object} $\tau_{\rm coll}$ of the same order as the cluster age
(for the Fig.~1, for $\tau_{\rm coll}$, an average object is assumed to have mass of $0.5 M_\odot$).
Roughly, the binaries with $\tau_{\rm coll}$ more than few Gyr, are not perturbed by dynamical encounters,
and with $\tau_{\rm coll}$ less than few Gyr -- only the hard binaries -- did undergo a dynamical encounter but survived.

Clearly  the efficiency of the destruction of binaries in realistic dense stellar systems is somewhere in between
these limiting examples, and also strongly affected by the evaporation 
of both binaries and single stars from the core due to high post-encounter velocities or supernova kicks,
and by continuing segregation of binaries and single stars from less dense halo,
as will be discussed below.

\begin{figure}
  \includegraphics[height=.35\textheight]{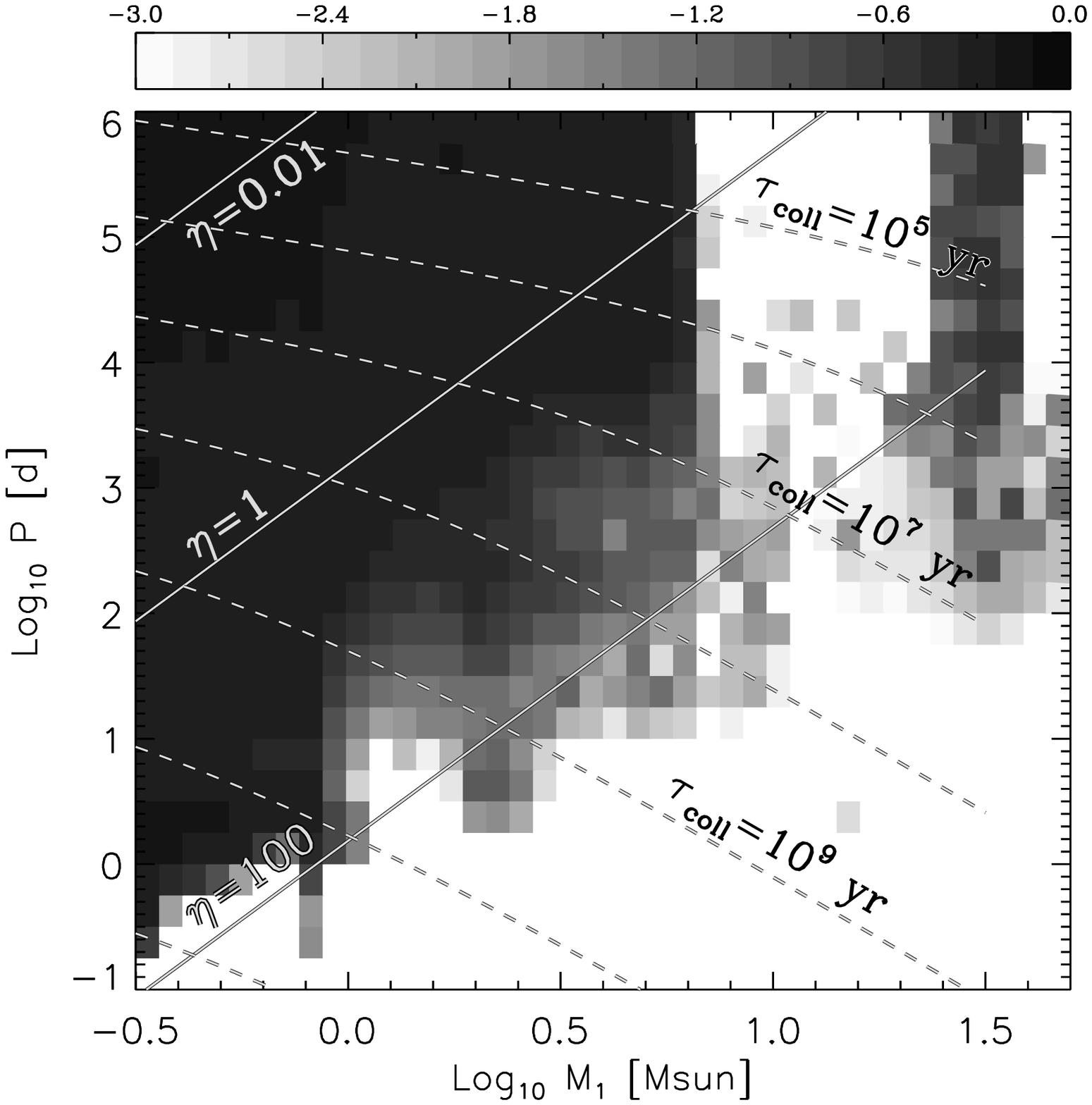}
  \includegraphics[height=.35\textheight]{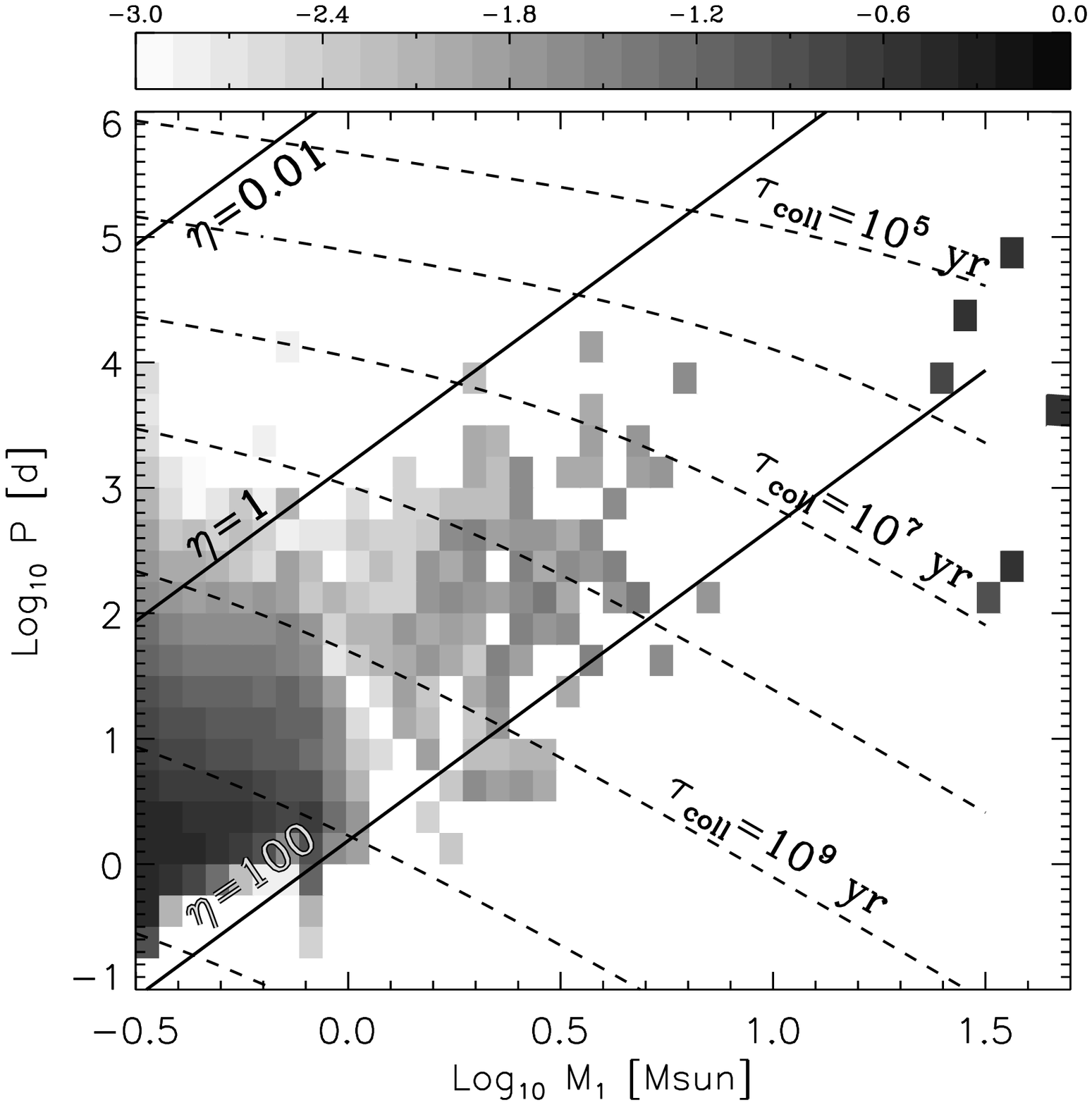}
  \caption{Probability density for a primordial binary to survive 10 Gyr as a binary: 
the effect of the evolution only (left), the effect of dynamics (right). Solid lines shows $\eta$, 
binary hardness, and dashed lines show $\tau_{\rm coll}$, collision time.}
\end{figure}

\subsection{Evolution of hard binaries}

Two main approaches currently used to model the dynamics of a dense stellar system are direct $N$-body, 
the most widely known realization is the family of $NBODY$ codes written by 
S.~Aarseth\footnote{See http://www.ast.cam.ac.uk/$\sim$sverre/web/pages/nbody.htm}, 
with the most recent version $NBODY6$;
and Monte Carlo (MC) methods, where encounters itself are usually treated with a
direct few-body integrator  (e.g., \citep{2001A&A...375..711F,2006MNRAS.371..484G}). 
MC methods are more capable of treating large populations and have been already used
to make models of observed globular clusters 
(e.g., M4 \citep{2008MNRAS.389.1858H}, NGC 6397 \citep{2009MNRAS.395.1173G}, 47 Tuc \citep{2010arXiv1008.3048G}),
while direct N-body is usually used to model open clusters 
(e.g. NGC 188 \citep{2010IAUS..266..258G}, M76 \citep{2005MNRAS.363..293H}),
but is starting to be also applied to globular cluster (e.g. NGC 6254 \citep{2010ApJ...713..194B}).
Here, we will refer MC to the specific realizations that
self-consistently model the global evolution of a cluster, using
MC techniques to sample the stellar distribution function when
applying the effects of two-body relaxation \citep{2009ApJ707.1533F}.
This not to be confused  with a {\it simplified} Monte Carlo method with a two-zone model of 
a cluster as in \cite{2005MNRAS.358..572I}.
In all the cases, the dynamical codes assume as well the use of 
a (binary) stellar evolution (population synthesis) code of a different complexity.

Due to still persisting computational power limitation, even with nowadays super-computers, including those 
equipped with GPUs, the numerical $N$-body models of globular clusters
are traditionally calculated either using small number of primordial binaries 
or small mass clusters (up to 100,000 stars,
e.g. \citep{2010MNRAS.tmp.1390H}), or rather very sparse.
The limitation so far comes not from the stellar evolution, which 
is either parametrized or tabulated, but from dynamics.
The maximum initial binary fraction used for a direct $N$-body modeling 
is 50\% for an (open) cluster made of 12,000 single and 12,000 hard binary stars and with core density
$n_{\rm c}=100-350$ pc$^{-3}$ \citep{2007ApJ...665..707H}. 
For relatively more massive clusters, up to 100,000 stars (a factor of 10 short compared
to observed massive and dense clusters) initial binary fraction  is usually 5\%.
(e.g. \citep{2010MNRAS.tmp.1390H}).

With the same initial conditions for hard binaries (low or medium density clusters, binary fraction $\le 50\%$ ), 
MC and direct $N$-body  do agree each other qualitatively and quantitatively: the core binary fraction increases with time
(see the comparison made in \citep{2009ApJ707.1533F}).
For larger initial binary fractions, MC shows the same behavior for almost all initial binary fractions
up to  $\phi_{\rm B}=90\%$ (here, all primordial binaries are hard).
The simulations were performed for cluster models with 
initially $10^5$ objects and with IMF extends from 0.15 to only 18.5 $M_\odot$ (MC is not designed to handle subsystems of rare
objects, like black holes (BHs), which could be produce by stars more massive than $18.5\, M_\odot$);
the highest initial core density $10^{4.5}$ pc$^{-3}$.

Interesting enough to note that the increase of the core binary fraction with time was due
to the two reasons. The first was theoretically expected to play the most significant role -- the segregation of binaries
from the halo to the core. However, especially in the case with binary fraction no being close to one,
the main net effect for the binary increase was at large due single stars evaporation from the core (in fact,
just a few \% of original core single stars remain in the core, the rest is evaporate either to the the clusters
outskirts, or lost from the cluster completely, then for binaries about a half of original hard binaries 
remains in the core). By the end of the core mass was decreased by at least an order
of magnitude with time.

\subsection{Comparison with observations: is there a steady binary-burning phase?}

In cases when MC models entered the binary-burning phase (not a common results among all models) before a Hubble time,
it was found that the core binary fraction in this phase steadily decreases with time. 
This behavior is consistent with the results of simplified two-zone Monte Carlo model described in \citep{2005MNRAS.358..572I},
where the binary-burning phase was imposed via adopted constant with time  cluster core properties.
It is likely then that efficient the hard binary fraction depletion can proceed 
only in a post-collapsed cluster. How common can this be?

As it was mentioned, in most of numerical simulations with hard binaries, the 
steady binary-burning regime is not observed. A cluster evolution towards the core collapse
can take up to a Hubble time. One other hand, the theoretical predictions for a cluster 
in a binary-burning phase do not match well with observations.
On Fig.~2 we show the comparison of the cluster cores predicted by theory for clusters in a binary-burning mode
and of the observed cluster cores. 
It can be seen, that the theoretical cluster cores are significantly
(by an order of magnitude) are smaller than the observed ones.
This likely indicates that the most of globular clusters in Milky Way is not in a binary-burning phase \citep{johnf}.

\begin{figure}
  \includegraphics[height=.35\textheight]{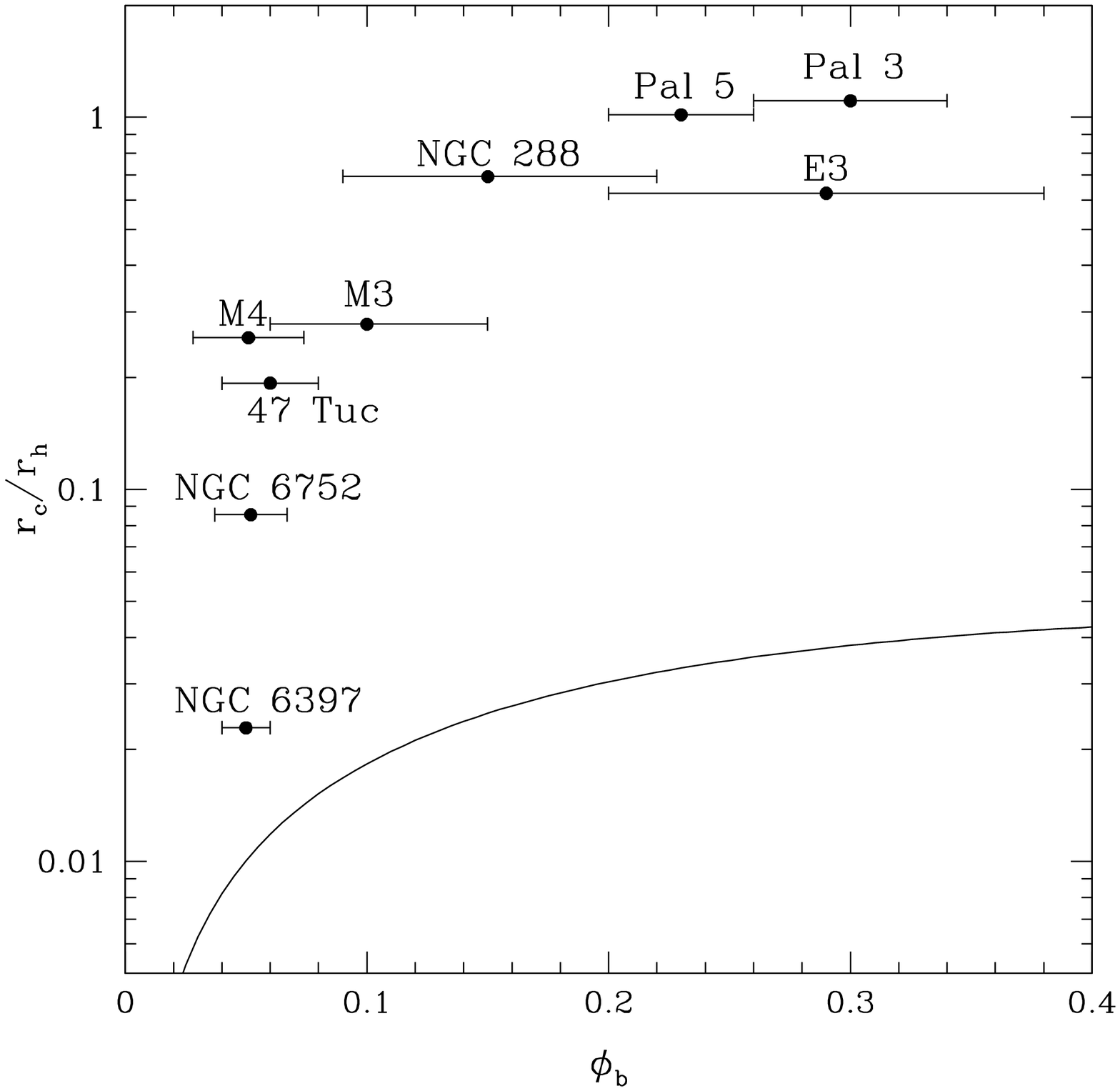}\includegraphics[height=.35\textheight]{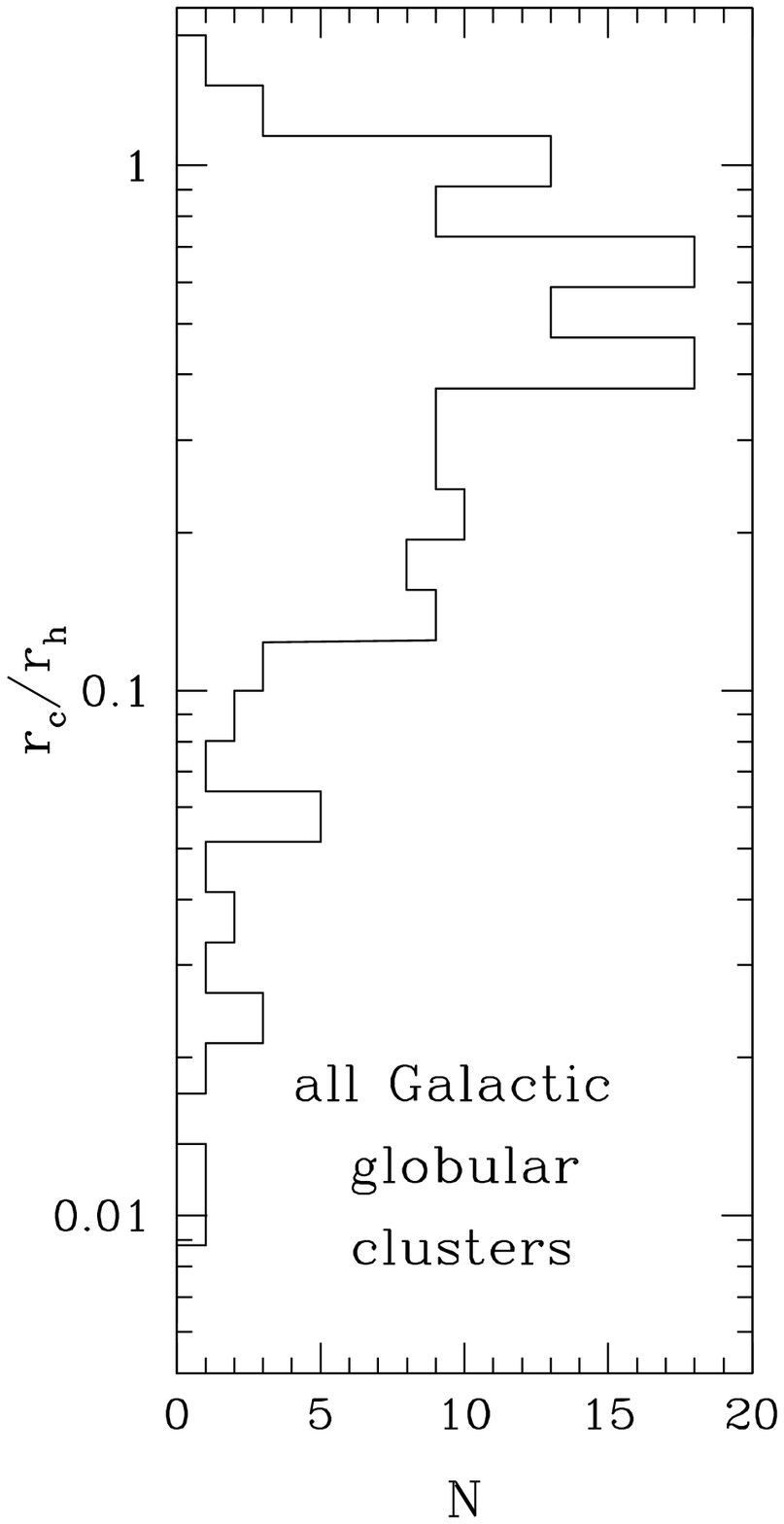} 
  \caption{The comparison of the theoretical relation for the cluster cores at binary-burning phase and the observed cluster cores;
 $\phi_{\rm B}$ is the core binary fraction. 
Credit for this figure is to John Fregeau, with thank to Craig Heinke for compiling the 
observational data with the same assumption (e.g., flat mass ratio) 
(\cite{2010ApJ...709.1183M, 2005MNRAS.358..572I, 2001ApJ...559.1060A,2004AJ....128.2274K,2004AJ....128.3019C,
1996ASPC...92..301V,2002AJ....123.1509B,2009AA...493..947S, 2005AJ....129.1934Z}). 
The theory curve comes from \citep{1994ApJ...431..231V}.  }
\end{figure}

\subsection{Soft binaries: can they still be neglected?}

The described above results were obtained using only initially hard binaries, as 
soft binaries have been traditionally ignored in both Monte Carlo
and direct $N$-body simulations. Mainly, this is because soft binaries
are expected to be destroyed on a very short time-scale and as such are not 
expected play a role on the globular cluster dynamics as a whole. 
The other reason, specific for $N$-body, is that each wide (soft) binary 
requires enormous calculation power, significantly slowing down the cluster's simulation.

However, largely unexpected result has been obtained in the model that included soft binaries,
and had an initial binary fraction $\phi_{\rm B}=90\% $.
In this case, after the rapid initial contraction of the core, the cluster
fairly quickly (compared to the models with only hard initial binaries, on the scale 
of just a few Myr) enters into a long-lived binary-burning phase keeping almost constant
binary fraction of 40\% \citep{2009ApJ707.1533F}.
The results were obtained for a cluster model with $5\times 10^5$ stars.

The physical reason for a different behavior of a cluster with a substantial  
contain of initial soft binaries is that this soft component acts, while present,  
as a significant cluster energy {\rm sink}. Soft binaries indeed are ionized 
quickly through encounters, absorbing clusters' energy.
As such, if the core was born with a significant number of soft binaries, 
it will rapidly contract on the time-scale of soft binaries destruction.

\subsection{What is a true binary fraction?} 

It is not widely recognized, but theoretically obtained binary fractions are  
very different from the observed ones. When describing simulations and its match 
to observed globular clusters, therefore, it is important to distinguish between a ``true'' theoretical binary fraction 
-- how many binary objects are present in a simulation,
independently on their characteristics, such as mass and luminosity, and
an ``observed'' binary fraction -- how many binaries would be found by an observer 
if this simulation would be a real cluster.

It can go either directions. For instance, when observers find binaries in a specific period range, 
an overestimation of binaries fraction can be done if it is assumed that the birth period distribution 
has been conserved for hard binaries, and only soft primordial binaries are destroyed.
In \citep{2005MNRAS.358..572I} it was shown that even hard binaries get destroyed,
when wider hard binaries are destroyed by dynamical encounters (a significant fraction of binaries up 
to $10kT$ is very vulnerable), and the evolution takes away very hard or intermediately hard but massive binaries (see also Fig.1). 
E.g., in \citep{2001ApJ...559.1060A}, the observationally derived binary fraction for 47 Tuc was found to be 13\%,
but if the change in periods distribution is taken into account, then the extrapolation from the observed binaries
can give only the binary fraction of about 6\%.

On the other hand, \citep{2009ApJ707.1533F} had demonstrated that even in simulations with only hard binaries, when
the overall ``true'' binary fraction in the core increases, the ``observed'' binary fraction {\it decreases}
for all simulations with the initial hard binary fraction above 40\%. For instance, a medium density cluster with
initial core density $10^{3.5}$ pc$^{-3}$ and initial binary fraction 90\%, evolved to only 30\% of observable binaries in the core
before the clusters was tidally disrupted. The ``true'' binary fraction though in this case approached almost 100\%.
``Hiding'' binaries here include binaries where two main sequence (MS) stars with a low-mass ratio would blend for an observer 
into one MS star, or binaries with one or two dim compact object as a companion.

\section{Binaries evolution: formation of compact binaries}

Although most dynamical interactions in dense cluster cores tend to destroy
binaries, some can form binaries from singles stars, and many modify them.
The most important processes are those that lead to a formation of
binaries both have a compact companion and are compact enough to start mass transfer.
Such binaries can be detected not only in our Milky Way, but in globular clusters in
distant galaxies as well, giving us a link between the statistical properties of
internal parameter of globular clusters and the efficiency of dynamical encounters.
In particular, low-mass X-ray binaries (LMXBs) are formed 
in globular clusters at the rate of 100 times exceeding that of in the field, 
per stellar mass unit \citep{1975ApJ...199L.143C}.
The most of binary millisecond pulsars (bMSPs) -- the likely termini of LMXBs evolution -- detected so far
are located in globular clusters \citep{2008AIPC..983..415R,2008IAUS..246..291R, 2010AAS...21545330L}.
It was even proposed that {\bf all} LMXBs and bMSPs are formed in globular clusters \citep{2007AIPC..924..649D}.
The importance of dynamical encounters for LMXBs formation seems to be well established, as the number of LMXBs 
was found to correlate well with the cluster dynamical properties \citep{2003ApJ...591L.131P} for all non-core-collapsed 
clusters. \citep{2008ApJ...673L..25F} explained the exception of core-collapsed clusters 
by our misunderstanding of clusters current dynamical states
and suggested that most globular clusters are still in the process of core contraction.
More recent observations showed that the overabundance of LMXBs in core-collapsed clusters is statistically significant,
and the number of X-ray sources in such clusters is almost independent on the cluster's dynamical properties 
\citep{2010PNAS..107.7164P}. The only LMXB with a BH and a white dwarf (WD) was detected in a globular cluster \citep{2007Natur.445..183M},
although their theoretically predicted numbers in the field should vastly exceed the number of LMXBs with a MS companion
\citep{2002MNRAS.329..897H, 2006ApJ...636..985I}, and 17 LMXBs with a MS companion in the Milky Way are observed.

There are several processes that lead to close binary formation: 
(1) binary companion exchange;  (2)  physical collision; (3)  tidal capture and (4) three-body binary formation.
The evolution of a dynamically formed binary  can be further 
perturbed by dynamical encounters:  a binary can be hardened and its eccentricity can be ``pumped''
 via multiple non-strong encounters;  it also can become a member of a triple.
The top channel to form a binary with a neutron star 
(NS) or a WD is a binary companion exchange, if binary fraction is at least few \% 
\citep{2006MNRAS.372.1043I,2008MNRAS.386..553I}.
A tidal capture (TC) operates only within a narrow range of periastra during a single-single star encounter 
 \citep{2006MNRAS.372.1043I};  as such, this channel does not account for more than a few \% of all formed binaries
with a compact companion. It has been argued that TCs could play an important role in the formation of 
ultra-luminous X-ray sources, with an intermediate-mass BH (IMBH) 
acquiring a companion through a TC \citep{2004ApJ...604L.101H, 2006MNRAS.372..467B}.
However, in most of TCs events with an IMBH, the energy dissipation rate  
in a captured star greatly exceeds the star's Eddington luminosity\citep{2006MNRAS.372..467B}. 
The fate of such a binary depends then on where inside the star the energy was deposited, 
and could be a merger \citep{1996MNRAS.279.1104P}.
Three-body binary formation is capable of creating a hard binary, but the formation  rate 
of binaries which are hard enough to become X-ray sources is small, even for
massive black holes \citep{2010ApJ...717..948I}.
Hardening, through multiple encounters, can shrink a hard binary towards it Roche lobe overflow, though a fraction
of the binaries that can successfully survive this path is not very large \citep{2010ApJ...717..948I}.
In the following we will review in more detail the processes that received attention in recent years:
formation via physical collisions and effect of the triples formation on compact binaries. 

\subsection{Formation of close binaries via physical collisions}

The formation of binaries via a physical collision between a red giant and
a compact object has been proposed first by \citep{1987ApJ...312L..23V}, in order to explain
ultra-compact X-ray binaries (UCXBs) formation. In this scenario, a collision leads to the formation
of a bound system that later might experience a common envelope and form a tight binary  (note
though that for low-mass giant remaining in globular clusters at current age, mass ratio favors 
a dynamically stable mass transfer). Smooth particle hydrodynamics (SPH) simulations showed later that
the stellar envelope can be disrupted or fully removed in close encounters,
with an eccentric binary formed as a result \citep{1991ApJ...377..559R, 1992ApJ...401..246D}.
Later, considering subgiants and early giants obtained using a stellar evolution code,
it was shown that (SPH) physical collisions between them and NSs,  
with small impact parameter $r_{\rm P}\le 1.5 R_{\rm RG}$ where $R_{\rm RG}$ is a giant radius, always 
lead to the complete expulsion of the envelope during a collision, resulting in
a tight and eccentric NS-WD binary; this binary shortly thereafter
decays its orbit and starts mass transfer \citep{2005ApJ...621L.109I, Lombardi06}. 
Formation rate of UCXBs by these encounters is sufficient to explain
the observed number of UCXBs in Galactic globular clusters or LMXBs in globular clusters near 
other galaxies \citep{2005ApJ...621L.109I}. The formation rates for UCXBs, being consistent with
the observed number, at the same time predict the number of bMSPs well exceeding the observed number
of bMSPs in globular clusters, and it was suggested that this type of LMXBs does not produce 
a radio bMSP \citep{2008MNRAS.386..553I}.

When the first BH-WD LMXBs was detected, 
it was plausible to consider a similar formation mechanism.
Detailed SPH numerical simulation showed that even though in most of encounters with  $r_{\rm P}$ up to $\sim 5 R_{\rm RG}$
a  bound system is formed, only a small fraction of them (with $r_{\rm P}$ up to $\sim R_{\rm RG}$) will 
lead to a formation of binaries that are tight enough and eccentric enough to start mass transfer in isolation \citep{2010ApJ...717..948I}.
A formation rate of BH-WD LMXBs through only physical collisions is about order of magnitude lower then required 
to explain the observed formation rate; the binary exchange channel does not provide binaries compact enough
to start the MT in isolation, so a sequence of post-collisional dynamical encounters is necessary. 
\subsection{Role of triples in compact binaries formation}
One of the outcomes of a binary-binary encounter is the formation of a  triple system 
(e.g. \citep{1975MNRAS.173..729H}), and some of the formed triples will be hierarchically-stable \citep{2001MNRAS.321..398M}.
In a typical dense cluster with central density $10^5$ pc$^{-3}$ and binary fraction of $\sim 10\%$, 
at 10 Gyr, about 5\% of all core binaries would have successfully formed a hierarchically stable triple during 1 Gyr
\citep{2008msah.conf..101I}; over whole life of the cluster, about half of binaries could
become a member of a triple. A typical formed triple has large ratio of orbital periods, $P_{\rm out}/P_{\rm in} \sim 1000$
and very large outer eccentricity, $e_{\rm out}=0.95\pm0.05$. About a half of formed triples is hard.

The effect of triples formation on the evolution of binaries population in globular clusters 
has just started to be recognized over the last several years. 
Driving force behind the triples' effect on binaries is that, if a triple has a large enough inclination, its secular evolution
is affected by Kozai mechanism \citep{1962AJ.....67..591K}. This mechanism causes large variations in the eccentricity 
and inclination of the star orbits and could drive the inner binary of the triple system to Roche lobe overflow,
when as result of the starting mass transfer a binary either merges before the next interaction, 
or start stable mass transfer. If one component of the inner binary is a non-degenerate star,
tidal interactions can be important during the periastra, when Kozai-induced eccentricity is at its maximum.
Shrinkage of a binary by a combination of Kozai cycling and tidal friction (KCTF; 
\cite{1979A&A....77..145M,1998MNRAS.300..292K,2006Ap&SS.304...75E,2007ApJ...669.1298F})
can be responsible for production of short-period active stars, e.g. BY-Dra-type binaries 
\citep{2009ApJ...703.1760M}.  KCTF would operate in a cluster if formed triple's Kozai time-scale \citep{1997AJ....113.1915I} 
will be smaller than $\tau_{\rm coll}$. As triples formation provides rather uniform space distribution for
inclinations, about 1/3 of all the formed triples affected by Kozai mechanism.

Blue stragglers, which are exotic stars with masses up to three cluster's turn-off mass but presumably still burning hydrogen, 
are detected both in globular and open clusters.
They are believed to be either produced by collisions between stars in clusters or stable or unstable mass transfer between
the components of primordial short-period binaries. Using a simplified Monte Carlo method, it was found that a fraction
of blue stragglers which can be provided by KCTF is comparable to that from 
physical collisions and primordial binaries   \citep{2008msah.conf..101I}.
KCTF also has been proposed to have a dominant role in the formation of blue stragglers  in open clusters 
and in the field \citep{2009ApJ...697.1048P}.
Specifically, it could naturally explain the large binary fraction of long-period blue stragglers binaries; 
these results are based on an assumption of primordial population of triples 
rather than on the population of dynamically formed triples.

As a star evolves, the role of the triple in the formation  of a binary with such a star is increasing,
and the probability that an inner binary would have a compact star as a companion is higher than for 
overall binary population \citep{2008msah.conf..101I}.
E.g., if KCTF in a triple with an inner binary consisting of a WD and a MS star 
leads not to a merger but to a stable mass transfer, the number of formed and present 
cataclysmic variables can be increased by up to 50\%,
if compared to simulations where KCTF is not taken into account.
If KCTF is taken into account, but leads to a merger, then the number of CVs could be depleted by third.

In a 47 Tuc type cluster, 5\% of all binaries with a NS companion would become a member of a hierarchically stable 
triple in a Gyr; in more dense cluster, Ter 5 - type, 15\% of all binaries with a NS would be a member of a triple.
In simulations, it appear that $\sim50\%$ of all bMSPs were in triples at some point in their past 
\citep{2008MNRAS.386..553I}. 

In simulations of globular clusters, since there is as yet no ``triple'' population synthesis,
a dynamically formed triple is simply broken, and the inner binary is shrunk to
the minimum periastron it could have through simple Kozai cycle, if Kozai time scale is shorter than  $\tau_{\rm coll}$. 
So in current clusters simulations, the mass transfer could be induced by triple formation, 
but not forced by being continuously in a triple.

There is, however, an UCXB 4U 1820-303, located in the globular cluster NGC 6624, with 
the orbital period $P$ of 685s \citep{1987ApJ...312L..17S, 1997ApJ...482L..69A}.
The stability value $\dot P/P=-(3.5\pm1.5)10^{-8}$ yr$^{-1}$ 
makes certain that 685s is the orbital period \citep{2001ApJ...563..934C}.
Secondary star is a He WD $0.06-0.08\,M_\odot$ \citep{1987ApJ...322..842R}.
Note that standard scenario of  Roche lobe mass transfer for a such
otherwise rather usual UCXB binary implies positive period derivative.
\citep{2001ApJ...563..934C} found that, in addition, 4U 1820-303 has the 
luminosity variation by a factor of 2 at a super-orbital period $P_{\rm so}\sim171$d.
It has been suggest that, since X-ray busts take place only at the flux minimum, 
the observed variability is due to intrinsic luminosity/accretion rate changes 
and not obscuration or changes of the projected area of the source due to precession.
The ratio between super-orbital and orbital periods (~22000) is too high for any kind 
of the disk precession at the mass ratio of the system \citep{1998MNRAS.299L..32L, 1999MNRAS.308..207W}.

It has been suggested that this binary is part of a hierarchical triple \citep{2001ApJ...563..934C},
with the third body having a mass $< 0.5M\, M_\odot$ (based on the lack of its optical detection)
and with the third body orbital period $P_{\rm out}\sim1.1$d (as eccentricity modulation is
expected to be on $P_{so}\simeq P_{\rm out}^2/P$).
\citep{Prodan10} studied this system in detail with the inclusion of the following:
perturbation from a third body on a longer period orbit; 
the quadrupolar distortion of stars due to their intrinsic spins and the further quadrupolar distortion due to their mutual gravity;
tidal friction in the equilibrium tide approximation; general relativity; mass transfer and gravitational radiation.
They found that the long period naturally arises if the system is librating around the stable fixed point in a
Kozai resonance. The observed system including the long period for eccentricity modulation can be reproduced 
with a NS with mass of $1.4\,M_\odot$, WD mass of $0.067M\,_\odot$ and third body mass 
of $0.55\,M_\odot$. The semi-major axes of the inner binary
is $a=1.32\times10^{10}$cm,  of the outer binary $a_{\rm out}=6.52a$ and corresponding period of 
$P_{\rm out} = 0.11$d. Initial eccentricity $e_0 = 0.002$,  initial outer binary eccentricity 
$e_{\rm out,0}=0.0001$ and initial inclination $65.66^{\rm o}$.
The amplitude of the eccentricity oscillations was found 
to be of $\sim 3\times 10^{-3}$, which is sufficient to enhance mass transfer enough to produce the observed
luminosity oscillations by factor of 2 \citep{2007MNRAS.377.1006Z}.

Although there is not yet a detected triple with a BH as a companion of the inner binary,
the role of dynamical triple formation for compact binaries with a BH could be even higher than for
compact binaries with a NS companion. For binaries with similar masses and orbital separation $a_1$
and $a_2$, the fraction of binary-binary encounters that result in a hierarchically stable triple formation is about 25\%
for all encounters that have a periastron within $\sim 7(a_1 + a_2)$ (for details see \citep{1983MNRAS.203.1107M}).
For binaries with high mass ratio where a more massive binary also is more compact,
this fraction approaches $30\%$ for encounters with a periastron within $\sim 20(a_1 + a_2)$,
essentially 3.5 times more efficient than equal mass binaries.
For binary fractions of a few per cent, triple formation for a binary with a BH is 
more frequent than strong encounters with single stars \citep{2010ApJ...717..948I}.
It has been shown that for a seed BH-WD binary with an initial separation of $a\le 80 R_\odot$, the
triple formation with subsequent Kozai cycle
coupled with gravitational wave emission will result in an LMXB formation during several Gyr.

\section{Conclusions}

In studies of binary fractions,  
it is crucial to distinguish between the ``observable'' and the ``true'' binary
fractions, they have different values and have different evolution. Soft binaries can
not be ignored in simulations: they act as an energy sink and could be the key in
achieving steady binary-burning phase. Triples play very important role in the formation
of compact binaries, specifically, X-ray binaries with a NS or a BH accretor.

\begin{theacknowledgments}
  NI acknowledges support from NSERC and Canada Research Chairs Program.
\end{theacknowledgments}



\bibliographystyle{aipproc}   

\bibliography{bin_gc}

\end{document}